\RequirePackage[normalem]{ulem} 
\RequirePackage{color}\definecolor{RED}{rgb}{1,0,0}\definecolor{BLUE}{rgb}{0,0,1} 
\RequirePackage{listings} 
\RequirePackage{color} 
\lstdefinelanguage{DIFcode}{ 
  moredelim=[il][\color{red}\sout]{\%DIF\ <\ }, 
  moredelim=[il][\color{blue}\uwave]{\%DIF\ >\ } 
} 
\lstdefinestyle{DIFverbatimstyle}{ 
	language=DIFcode, 
	basicstyle=\ttfamily, 
	columns=fullflexible, 
	keepspaces=true 
} 
\lstnewenvironment{DIFverbatim}{\lstset{style=DIFverbatimstyle}}{} 
\lstnewenvironment{DIFverbatim*}{\lstset{style=DIFverbatimstyle,showspaces=true}}{} 

\documentclass[sigconf]{acmart}
\AtBeginDocument{%
  }

\usepackage{booktabs}   
\usepackage{array}
\usepackage{tabularx}   
\usepackage{multirow}   
\usepackage{rotating}   
\usepackage{longtable}  
\usepackage{caption}    
\usepackage{enumitem}
\usepackage[utf8]{inputenc}

\copyrightyear{2026}
\acmYear{2026}
\setcopyright{cc}
\setcctype{by}
\acmConference[CHI EA '26]{Extended Abstracts of the 2026 CHI Conference on Human Factors in Computing Systems}{April 13--17, 2026}{Barcelona, Spain}
\acmBooktitle{Extended Abstracts of the 2026 CHI Conference on Human Factors in Computing Systems (CHI EA '26), April 13--17, 2026, Barcelona, Spain}
\acmDOI{10.1145/3772363.3798409}
\acmISBN{979-8-4007-2281-3/2026/04}

\usepackage{tabu}                      
\usepackage{booktabs}                  
\usepackage{lipsum}                    
\usepackage{mwe}                       

\usepackage{enumitem}
\usepackage{soul}
\usepackage{listings}
\usepackage{quoting}

\newcommand{\bpstart}[1]{
\noindent{\textbf{#1}}%
}

\newcommand{\app}{SuperProvenanceWidgets\xspace}
\newcommand{\apps}{SuperProvenanceWidgets\xspace}

\newcommand{\paragraphHeadingSpace}{\vspace{4px}}

\usepackage{setspace} 
\lstdefinelanguage{TypeScript}{
    basicstyle=\small\ttfamily\color{blue}
}

\lstdefinelanguage{HTML5}{
    language=HTML,
    basicstyle=\small\ttfamily\color{magenta}
}

\lstdefinelanguage{CSS}{
    basicstyle=\small\ttfamily\color{orange}
}

\lstdefinelanguage{Angular}{
    language=HTML,
    basicstyle=\small\ttfamily\color{magenta},
    alsoletter={<>=-},
    morekeywords={
        provenance-,
        slider,
        dropdown,
        multiselect,
        radiobutton,
        checkbox,
        inputtext,
        provenance-slider,
        provenance-inputtext,
        provenance-dropdown,
        provenance-multiselect,
        provenance-radiobutton,
        provenance-checkbox
    },
    deletekeywords={value, selected, data}
}

\lstdefinelanguage{TS}{
    language=Java,
    morekeywords={
        SliderProvenance,
        InputTextProvenance,
        Provenance,
        "interaction",
        "time",
        number,
        string,
        ChangeContext,
        Options,
        Option,
    },
    deletekeywords={label, true, if, return}
}

\lstdefinestyle{TypeScript}{
  language=TypeScript,
  basicstyle=\small\ttfamily,
  keywordstyle=\color{blue},
  commentstyle=\color{green!60!black},
  stringstyle=\color{red},
  breaklines=true,
  showstringspaces=false,
  numbers=left,
  numberstyle=\tiny,
  frame=lines,
  captionpos=b
}

\lstdefinestyle{HTML}{
  language=HTML5,
  basicstyle=\small\ttfamily,
  keywordstyle=\color{blue},
  commentstyle=\color{green!60!black},
  stringstyle=\color{red},
  breaklines=true,
  showstringspaces=false,
  numbers=left,
  numberstyle=\tiny,
  frame=lines,
  captionpos=b
}

\lstdefinestyle{Angular}{
  language=Angular,
  basicstyle=\small\ttfamily,
  keywordstyle=\color{blue},
  commentstyle=\color{darkgray},
  stringstyle=\color{teal},
  breaklines=true,
  showstringspaces=false,
  numbers=left,
  numberstyle=\tiny,
  frame=topline,
  captionpos=b,
}

\lstdefinestyle{TS}{
  language=TS,
  basicstyle=\small\ttfamily,
  keywordstyle=\color{teal},
  commentstyle=\color{lightgray},
  stringstyle=\color{teal},
  breaklines=true,
  showstringspaces=false,
  numbers=left,
  numberstyle=\tiny,
  frame=none,
  belowskip=0em,
  captionpos=b
}

\lstdefinestyle{JSX}{
  language=JSX,
  basicstyle=\small\ttfamily,
  keywordstyle=\color{teal},
  commentstyle=\color{lightgray},
  stringstyle=\color{teal},
  breaklines=true,
  showstringspaces=false,
  numbers=left,
  numberstyle=\tiny,
  frame=none,
  belowskip=0em,
  captionpos=b
}

\newcommand{\checkboxicon}[1]{\includegraphics[width=2em]{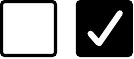}}

\newcommand{\radiobuttonicon}[1]{\includegraphics[width=2em]{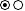}}

\newcommand{\slidericon}[1]{\includegraphics[width=3em]{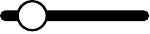}}

\newcommand{\rangeslidericon}[1]{\includegraphics[width=3em]{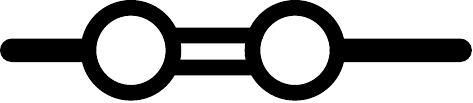}}

\newcommand{\singleselecticon}[1]{\includegraphics[width=3em]{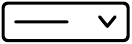}}

\newcommand{\multiselecticon}[1]{\includegraphics[width=3em]{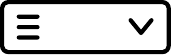}}

\newcommand{\inputtexticon}[1]{\includegraphics[width=3em]{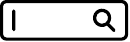}}

\newcommand{\aggmodeicon}[1]{\includegraphics[width=10pt]{figures/logos/logo-aggregate-rotated.pdf}}
\newcommand{\tempmodeicon}[1]{\includegraphics[width=10pt]{figures/logos/logo-temporal-rotated.pdf}}
\newcommand{\disabledmodeicon}[1]{\includegraphics[width=10pt]{figures/logos/logo-disabled-rotated.pdf}}

\newcommand{\coloredborderprovbutton}[1]{\includegraphics[width=14pt]{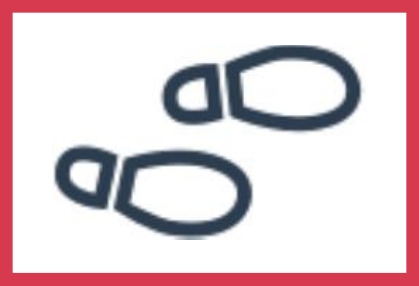}}
\newcommand{\coloredbgprovbutton}[1]{\includegraphics[width=14pt]{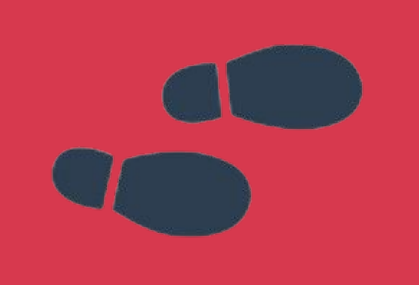}}

\newcommand{\cut}[1]{}

\begin{document}

\title{SuperProvenanceWidgets: Tracking and Visualizing Analytic Provenance \emph{Across} UI Control Elements}

\author{Antariksh Verma}
\email{averma@connect.ust.hk}
\orcid{0009-0007-2267-4363}
\affiliation{%
  \institution{The Hong Kong University of Science and Technology}
  \state{Hong Kong S.A.R.}
  \country{China}
}

\author{Kaustubh Odak}
\email{kaustubhodak1@gmail.com}
\orcid{0009-0006-5873-7891}
\affiliation{%
  \institution{Amazon Web Services}
  \city{Seattle}
  \state{Washington}
  \country{United States}
}

\author{Arpit Narechania}
\email{arpit@ust.hk}
\orcid{0000-0001-6980-3686}
\affiliation{%
  \institution{The Hong Kong University of Science and Technology}
  \state{Hong Kong S.A.R.}
  \country{China}
}

\renewcommand{\shortauthors}{Verma, Odak and Narechania}

\begin{abstract}
ProvenanceWidgets is an existing JavaScript library that tracks the recency and frequency of user interactions with individual UI controls (e.g., range sliders and dropdowns) and dynamically overlays this provenance onto them. In this work, we introduce \textbf{SuperProvenanceWidgets}, an extension to ProvenanceWidgets featuring a new SuperWidget that similarly tracks and visualizes provenance but across multiple UI controls, enabling users to understand how, when, and whether different UI controls were used. Through three example usage scenarios, we demonstrate how this cross-control SuperWidget helps (a) audit and share analysis workflows, (b) surface and mitigate exploration biases, and (c) facilitate user interface design and personalization. We also perform a technical self-assessment using the Cognitive Dimensions of Notations to evaluate the library's usability for developers. SuperProvenanceWidgets is integrated into the ProvenanceWidgets library and is available as open-source software at \textbf{\href{https://ProvenanceWidgets.github.io}{ProvenanceWidgets.github.io}}, empowering developers to build advanced provenance applications.
\end{abstract}



\begin{teaserfigure}
  \centering
  \includegraphics[width=0.95\linewidth]{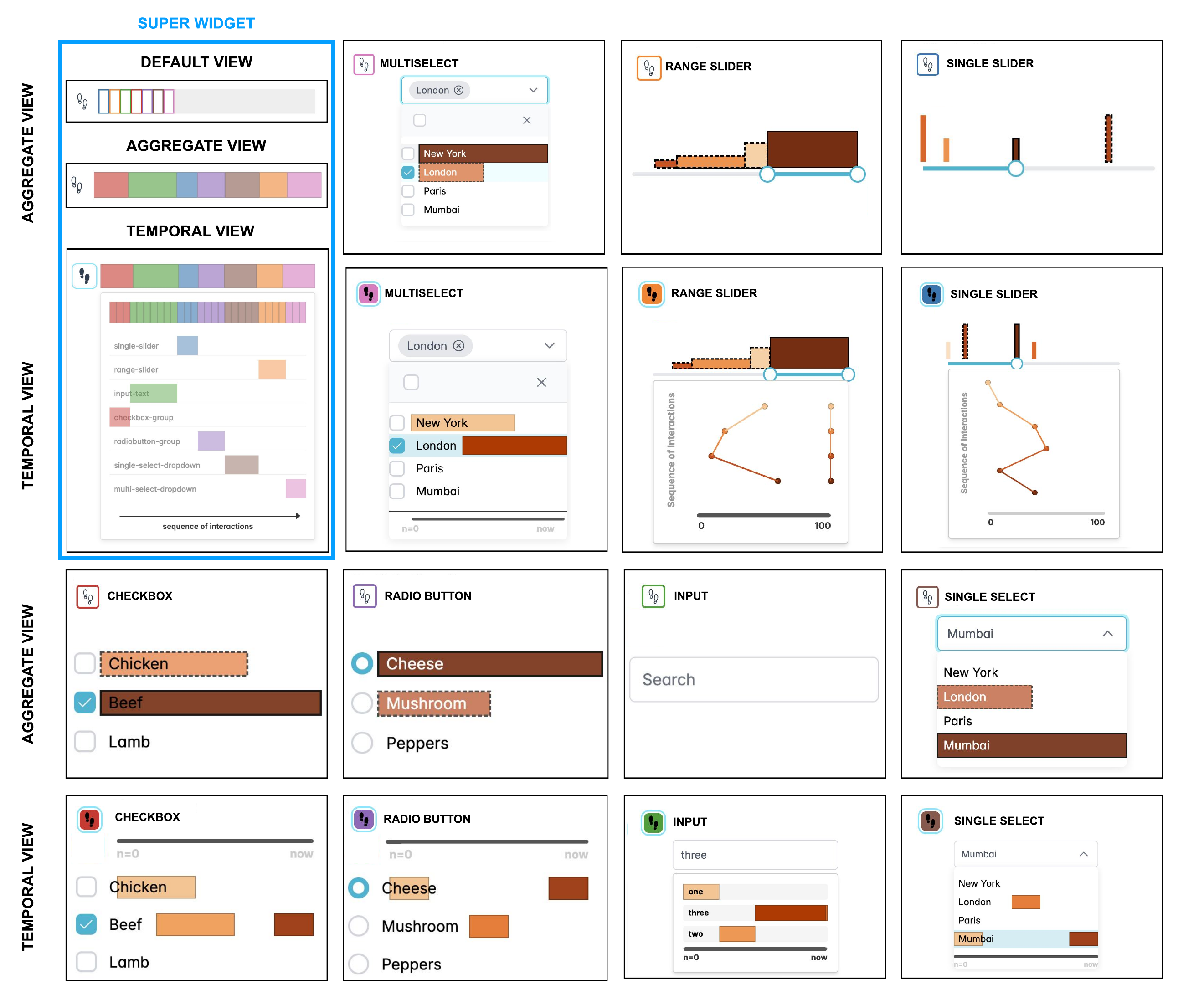}
  \caption{\apps: a JavaScript library comprising UI controls and a \emph{SuperWidget} showing an aggregated overview as well as a detailed temporal history of interactions both within and across UI controls.}
  \Description{}
  \label{fig:interface}
\end{teaserfigure}

\maketitle

\section{Introduction and Background}

Analytic provenance is the documented history of how users engage with and manipulate data visualizations~\cite{pike2009scienceinteraction, north2011analytic,ragan2015characterizing}. 
Documenting this history is important because our memory has a limited capacity to remember and track our prior interactions with data. These limitations occur both in quantity and over time as memories decay~\cite{miller1994magical, liu2014effects}. This creates a barrier to effective data exploration.

Consequently, provenance has been used to address this barrier in multiple ways. For instance, it supports sensemaking~\cite{nguyen2016sensemap} and decision-making~\cite{madanagopal2019analytic} with data. It helps evaluate visualization systems~\cite{bylinskii2017learning, gomez2012modeling}, design adaptive systems~\cite{walch2019lightguider}, improve machine learning model performance~\cite{endert2012semantic}, replay and replicate analytic workflows~\cite{bavoil2005vistrails, shrinivasan2009connecting}, and generate summary reports~\cite{chen2010click2annotate, gratzl2016visual}.

Provenance has also been shown to drive deeper analytic insights. For instance, it facilitates unique data discoveries~\cite{feng2017hindsight, willett2007scented}, improves user confidence~\cite{block2023influence} and inspiration~\cite{dunne2012graphtrail} levels, and increases contextual awareness~\cite{skopik2005improving} and recall~\cite{dunne2012graphtrail} of previously visited data. In some cases, it even causes surprise~\cite{narechania2025provenancelens}.

In terms of tooling, existing tools support provenance tracking across diverse domains, including data analysis platforms~\cite{north2011analytic}, code editors~\cite{footstepsvscode}, computational notebooks~\cite{gadhave2024persist,2024_loops}, workflow systems~\cite{bavoil2005vistrails,revisit2023ding}, collaborative environments~\cite{ellkvist2008using,sarvghad2015exploiting,badam2017supporting}, websites~\cite{googleanalytics}, UI controls~\cite{narechania2024provenancewidgets}, and even video games~\cite{drachen2015behavioral, kohwalter2017capturing}.

Most relevant to our work are software libraries that enable developers to build custom provenance-aware tools with cross-platform consistency~\cite{aigner2013evalbench, cutler2020trrack, baudisch2006phosphor, willett2007scented, narechania2024provenancewidgets}. For example, Trrack~\cite{cutler2020trrack} is a JavaScript library for creating and tracking interaction histories in web applications, supporting action recovery, reproducibility, collaboration, and logging.
ProvenanceWidgets~\cite{narechania2024provenancewidgets} is another JavaScript library that tracks and dynamically overlays analytic provenance, in the form of the recency and frequency of user interactions, onto UI controls such as sliders and dropdowns. 
However, ProvenanceWidgets \emph{independently} visualizes provenance on \emph{individual UI controls}; there is no way to analyze the aggregated summary or the detailed temporal sequence and duration of interactions \emph{across multiple UI controls}. 
For instance, when a user combines an input field and range slider with a checkbox group to identify trends across variables, developers using ProvenanceWidgets or Trrack must manually extract logs from each control and analyze them separately. This workflow is labor-intensive and redundant.
So we ask: \emph{How can we (better) track and visualize provenance \emph{across} different UI controls?}

In response, we present \textbf{\apps}, an extension to ProvenanceWidgets that tracks and visualizes provenance \emph{across multiple UI controls}. This enables users to understand not just \emph{what} controls were used, but \emph{how}, \emph{when}, \emph{for how long}, and \emph{in what sequence}—revealing the temporal relationships between interactions that individual control tracking obscures.

We achieve this through two key advancements to the ProvenanceWidgets framework.
First, we introduce a new modular architecture based on the \textit{Separation of Concerns}~\cite{dijkstra1982}. We restructured ProvenanceWidgets by de-coupling it into independent modules—separating provenance capture, visualization, and cross-control aggregation. This gives developers fine-grained control to customize and extend the library for diverse use cases.
Second, we introduce \textit{SuperWidget}, a novel multi-widget control that aggregates and visualizes \emph{super-provenance} across controls. It reveals temporal sequences, interaction durations, and usage patterns that show how users orchestrate multiple controls during analysis.

We demonstrate the impact of these advancements through three usage scenarios: (a) auditing and sharing complex analysis workflows~\cite{callahan2006vistrails,cutler2020trrack}, (b) surfacing and mitigating human~\cite{wall2022lrg} and exploration biases~\cite{narechania2021lumos,paden2024biasbuzz}, and (c) enabling user interface design~\cite{siroker2015b} and personalization~\cite{walch2019lightguider}. 
We also evaluate the library's developer usability through the Cognitive Dimensions of Notations~\cite{blackwell2001cognitive}.
\app is integrated into ProvenanceWidgets and is available as open-source software at \textbf{\href{https://ProvenanceWidgets.github.io}{ProvenanceWidgets.github.io}}, empowering developers to build advanced provenance applications.
\section{\app}


\subsection{Design Goals}

Our design of \apps builds upon the goals of the original ProvenanceWidgets library, and is driven by five goals.

\begin{enumerate}
    \item [G1] \textbf{Cross-widget provenance tracking.} The library should keep track of provenance within as well as across UI controls.

    \item [G2] \textbf{Real-time visual overlays.} The library should visualize aggregated summaries as well as temporal histories of users' analytic provenance within and across multiple UI controls, in real time.
    
    \item [G3] \textbf{Control selection.} The library should enable users to intuitively identify and navigate to individual UI controls.
    
    \item [G4] \textbf{Action recovery.} The library should enable users to revert to previous states of the UI controls.

    \item [G5] \textbf{Modular design.} The library should be composed of independent, composable units to enable developers to customize, extend, and integrate only the components they need.
\end{enumerate}

\subsection{Design Process}

\begin{figure*}[h!]
    \centering
    \includegraphics[alt={Sankey flow diagram}, width=0.9\linewidth]{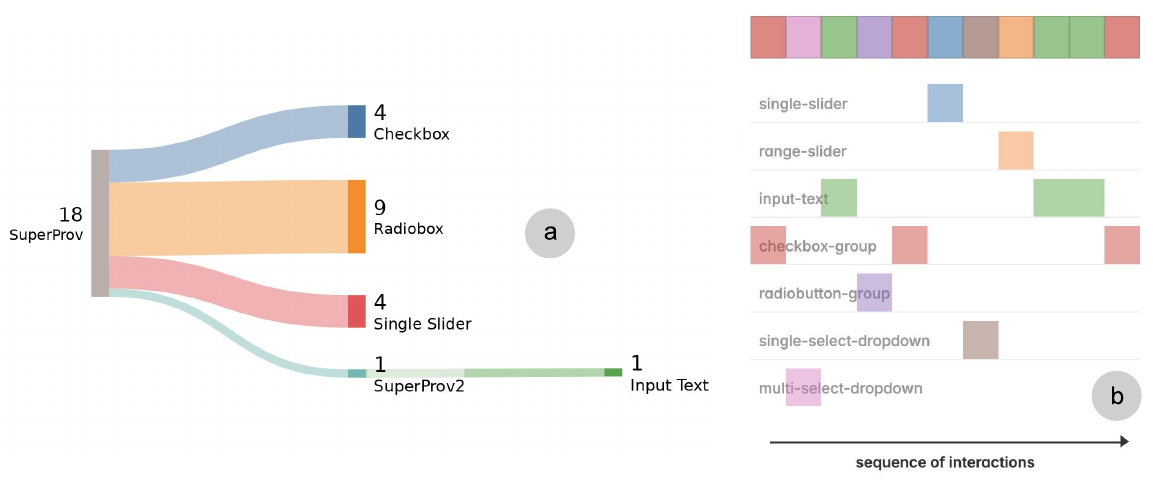}
    \caption{Different representations of user interactions (a) as a Sankey flow diagram and (b) as a Gantt chart.}
    \label{fig:external-viz}
\end{figure*}

\begin{figure}[h!]
    \centering
    \includegraphics[alt={Internal visualization}, width=0.7\linewidth]{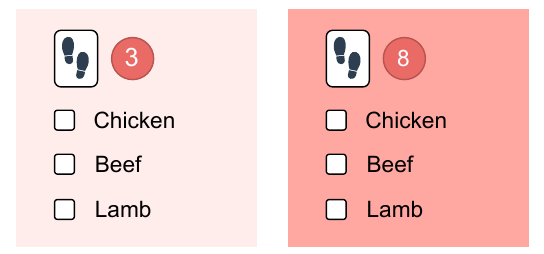}
    \caption{Two checkboxes with different interaction statuses and different visual properties.}
    \label{fig:internal-viz}
\end{figure}

To achieve our design goals, we systematically explored visualization strategies for presenting cross-widget provenance (\textbf{G1}). We considered both \textbf{ex-situ}--wherein cross-provenance is shown in a separate widget--and \textbf{in-situ}--wherein cross-provenance is shown on each UI control directly--approaches. All this, while balancing navigation support (G3) with minimal interference during analysis.

\subsubsection{Ex-situ Visualization.}
In ex-situ approaches, provenance data is displayed in a separate widget that users can access as needed. This approach also simplifies navigation to specific UI controls (\textbf{G3}). We prototyped two designs: \emph{Sankey diagram} and \emph{Gantt chart}. 

\paragraphHeadingSpace\bpstart{Sankey diagram.} Figure~\ref{fig:external-viz}a shows this design where the height of each band corresponds to number of interactions, highlighting the \textit{nestedness} of the flow. Sankey diagrams excel at revealing high-level patterns like which control combinations users favor most and how interactions branch across widgets. However, they obscure precise timing and make it difficult to recover exact UI states for replay.

\paragraphHeadingSpace\bpstart{Gantt chart.} Figure~\ref{fig:external-viz}b shows this design where the position of each bar illustrates the sequence of all interactions across widgets. Gantt charts clearly encode interaction timing, duration, and precise ordering, enabling action recovery by clicking any bar to restore that UI state. Their weakness is visual clutter when interaction volume is high, though this can be mitigated via filtering and zooming.

\subsubsection{In-situ Visualization.}
In in-situ approaches, we overlay provenance directly onto UI controls for immediate context. We explored two designs (Figure~\ref{fig:internal-viz}): \emph{widget background} and \emph{interaction badges}.

\paragraphHeadingSpace\bpstart{Widget background.} In this design, control backgrounds are shaded darker or lighter based on recency of last interaction. This provides instant ``at-a-glance'' feedback about which controls were most recently used without requiring extra screen space. However, it conveys no information about interaction frequency, duration, or sequence across multiple controls.

\paragraphHeadingSpace\bpstart{Interaction badges.} In this design, numeric badges beside controls display interaction counts. This reveals usage frequency at a glance and works well for comparing relative activity across controls. However, badges create persistent visual clutter during active analysis and cannot show temporal relationships or precise timing between interactions.

\paragraphHeadingSpace\noindent \textbf{We ultimately favored ex-situ Gantt charts} over other designs While in-situ encodings provide immediate context, their persistent overlays interfere with primary analysis tasks. Sankey diagrams reveal useful high-level patterns but obscure precise timing and state recovery, whereas Gantt charts explicitly encode interaction sequence and duration along a continuous timeline to fully support temporal tracking and action recovery (\textbf{G2}, \textbf{G4}). We integrated this Gantt chart into a new \emph{SuperWidget}, described next.

\subsection{SuperWidget: an extension to ProvenanceWidgets}
\label{ext-design}

\textbf{SuperProvenanceWidgets} extends ProvenanceWidgets~\cite{narechania2024provenancewidgets} by introducing the \emph{SuperWidget} for cross-control provenance (G1). It maintains full backward compatibility, supporting all original UI controls: radio buttons~\radiobuttonicon{}, checkboxes~\checkboxicon{}, single sliders~\slidericon{}, range sliders~\rangeslidericon{}, dropdowns~\singleselecticon{}, multiselects~\multiselecticon{}, and input text fields~\inputtexticon{}. The \emph{SuperWidget} comprises two views:

\begin{figure}[ht]
    \begin{minipage}{\linewidth}
        \centering
        \includegraphics[alt={Figure showing aggregate view}, width=0.95\linewidth]{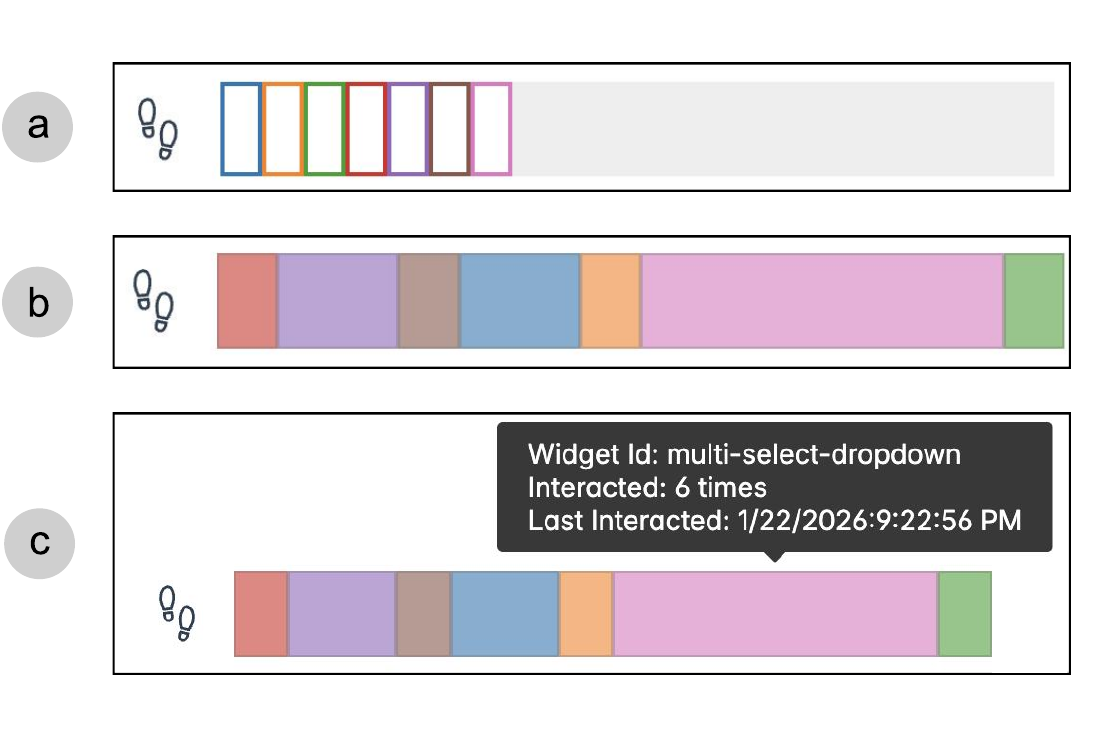}
        \caption{Aggregate View: (a) empty interaction status, (b) interacted with components, (c) hovering over colored boxes}
        \label{fig:aggregate-view}
    \end{minipage}\hfill
    \begin{minipage}{0.5\textwidth}
        \centering
        \includegraphics[alt={Figure showing aggregate view}, width=0.95\linewidth]{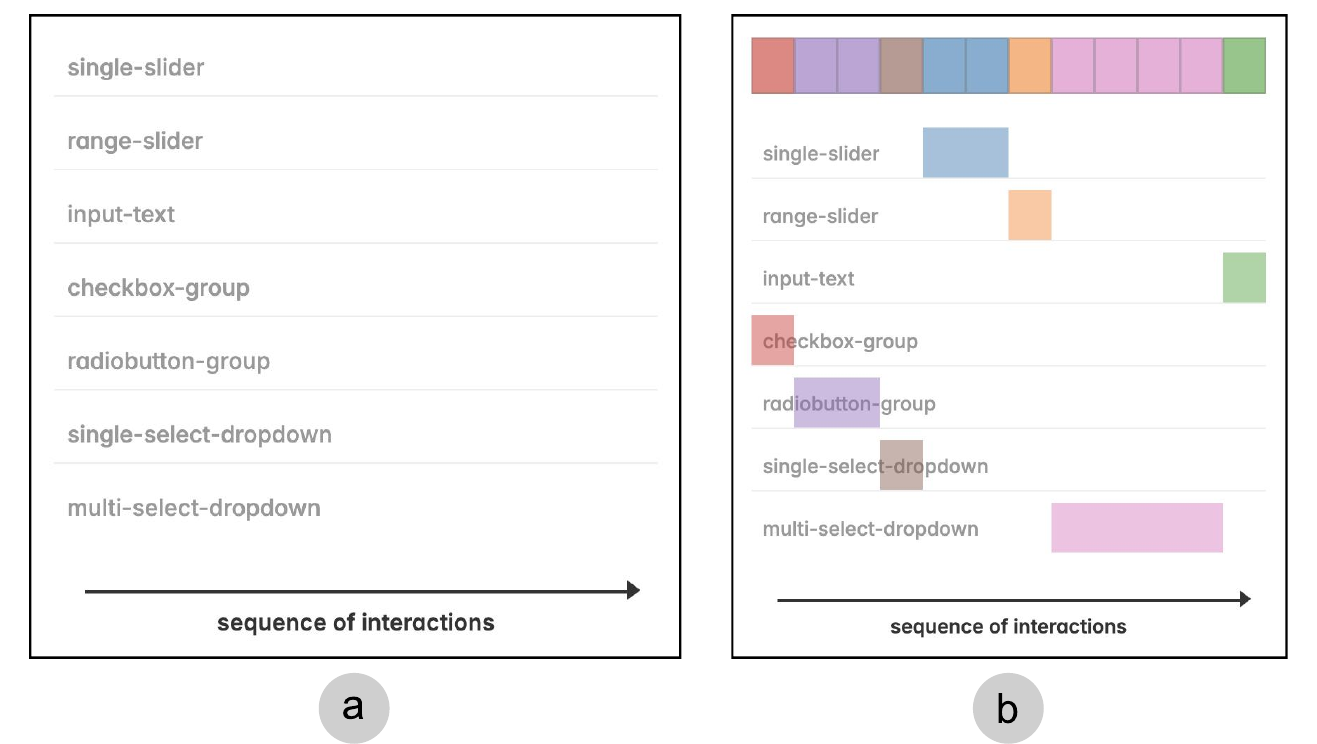}
        \caption{Temporal View: (a) empty interaction status, (b) interaction status after having interacted with multiple components}
        \label{fig:temporal-view}
    \end{minipage}
\end{figure}

\paragraphHeadingSpace\bpstart{Aggregate View.} 
To provide a compact overview of cross-widget provenance activity, we designed an Aggregate View (Figure~\ref{fig:aggregate-view}b). Each colored box (pink, green, red, etc.) represents one UI control (input box, checkbox group, radio group, single/multi-select dropdown, single/range slider). Colors are automatically assigned via the \emph{d3-scale-chromatic}\footnote{\url{https://d3js.org/d3-scale-chromatic}} JavaScript library for clear distinction.

Initially, unused controls show just a colored border with no fill (Figure~\ref{fig:aggregate-view}a). As interactions occur, boxes grow in size proportional to total interaction count and reposition based on recency. Hovering reveals details like widget ID, interaction count, and last interaction timestamp (Figure~\ref{fig:aggregate-view}c). Clicking navigates to the corresponding control (\textbf{G3}), bringing it into focus, while the overview supports real-time aggregated visualization (\textbf{G2}).

\paragraphHeadingSpace\bpstart{Temporal View.} To visualize the provenance across widgets over time, we designed an overlaid popover below the SuperWidget, showing full interaction history (Figure~\ref{fig:temporal-view}b). It lists all widgets by ID alongside a Gantt chart where the x-axis represents interaction sequence. Each row compiles that widget's interactions as colored boxes, revealing timing and ordering information.

To connect SuperWidget colors with individual UI controls, we enhanced ProvenanceWidgets' provenance button (with footprint icon) with matching colored borders~\coloredborderprovbutton{}. On click, the button fills solid with the same color~\coloredbgprovbutton{}, enabling instant visual identification between SuperWidget boxes and their corresponding controls.

\subsection{Architecture}

Existing provenance libraries like ProvenanceWidgets~\cite{narechania2024provenancewidgets} and others~\cite{camisetty2019enhancing} often use monolithic architectures that bundle all functionality together, resulting in unnecessarily large libraries that developers must include entirely. We redesigned ProvenanceWidgets based on the \textit{Separation of Concerns} principle~\cite{dijkstra1982}. This decomposes \textbf{SuperProvenanceWidgets} into three independent, composable modules that developers can mix and match (\textbf{G5}):

\begin{enumerate}
    \item \textbf{Compute Module.} The backend engine that captures user interactions, computes provenance statistics (recency, frequency, sequences), and stores data in memory for real-time access.
    
    \item \textbf{UI Module.} The frontend library providing pre-built SuperWidget components (Aggregate View, Temporal View) that connect to any Compute backend.
    
    \item \textbf{Scents Module.} Optional visualization extensions that overlay provenance cues (colored borders, badges) directly onto individual UI controls.
\end{enumerate}

Each module operates independently but integrates seamlessly via well-defined interfaces. Developers can use just the Compute module for custom UIs, combine UI+Compute for complete solutions, or add Scents for in-situ visualization--resulting in significantly smaller bundles and greater flexibility.

\subsection{Implementation}

The \textbf{Compute} module for \app is implemented in TypeScript, and the \textbf{UI} and \textbf{Scents} are implemented in React.js; these modules can be used as \emph{Web Components} making them framework-agnostic.

\subsubsection{User Interface API}

We declare our UI API below in terms of providers, hooks and components.

\begin{enumerate}
    \item \textbf{Providers.} Used to manage sharing of provenance-data across different components (\textbf{G1}).
    
    \begin{lstlisting}[style=TS]
<ProvenanceProvider>
    <App /> {/* Render root component */}
</ProvenanceProvider>
    \end{lstlisting}

    \vspace{0.1cm}
    
    Additionally, wrapping the application with the provider adds hierarchy, so developers can wrap the portion of the application that needs use of provenance, as opposed to making it available throughout the application.
    
    \item \textbf{Hooks.} Used to invoke getters and setters for provenance-data. The hook we expose is \verb|useProvenance|.

    \begin{lstlisting}[style=TS]
const [registeredComponents, setRegisteredComponents] = useProvenance()
    \end{lstlisting}

    \vspace{0.1cm}
    
    The hook returns two objects: \verb|registeredComponents| and \verb|setRegisteredComponents|. The former is an array that stores the global provenance of every single widget, whereas the latter is a provenance modifier function.
\end{enumerate}

\subsubsection{Scents API} The Scents API exposes a \verb|<Bars />| component to render scent bars, with the following properties.

\begin{enumerate}
    \item \textbf{guidance}: Provenance data object containing aggregate and domain information used to populate and scale the bars.
    \item \textbf{orientationScheme}: A function that maps normalized values to colors for the bars based on the color domain.
    \item \textbf{barKeys}: Position bars along the secondary dimension (horizontal/vertical).
    \item \textbf{encodings}: An object specifying the visualization layout -- orientation, positionDomain and colorDomain.
    \item \textbf{width} and \textbf{height}: SVG dimensions; dimensions are used for scaling bar lengths.
\end{enumerate}

\subsubsection{Compute API} The Compute API implements a single abstract \textit{GenericProvenance} class, and then extends it into the following classes to expose type-level functionality.

\begin{enumerate}
    \item \textbf{NumericProvenance}: Adds min/max bounds to track the range of numeric data.
    \item \textbf{RangedProvenance}: Handles ranged slider interactions storing paired [lowValue, highValue] data with count tracking and edge boundary management for visualization.
    \item \textbf{SelectionProvenance}: Tracks multi-select interactions recording which items were selected/unselected at each timestamp, maintaining selection and interaction counts per item.
    \item \textbf{TextProvenance}: A type alias of GenericProvenance specialized for string values, used for text input tracking.
    \item \textbf{SuperProvenance}: Aggregates multiple widget provenances (sliders, dropdowns, checkboxes, etc.), allowing registration of multiple widgets and tracking overall interaction counts.
\end{enumerate}

\section{Usage Scenarios}

We illustrate \textbf{\apps}' capabilities through three scenarios that demonstrate how cross-control provenance can support common analysis needs.

\subsection{Audit and share analysis workflows.}

Imagine a data science team tasked with auditing a colleague's complex analysis session to ensure reproducibility and validate key decisions. Without aggregated provenance, reviewers must sift through raw logs or replay sessions manually, which is time-consuming and error-prone, especially when interactions span dozens of UI controls like sliders, dropdowns, and checkboxes.

With SuperProvenanceWidgets, developers can connect custom data sources to the Compute module, which automatically records every interaction across controls. The SuperWidget's Aggregate View immediately reveals dominant controls through large, colored boxes sized by interaction frequency and positioned by recency, while the Temporal View displays precise sequences and durations via intuitive Gantt bars. Team members can easily export this structured provenance--complete with widget IDs, timestamps, and counts--for asynchronous review, quickly confirming appropriate use of critical controls or spotting anomalies for discussion.

\subsection{Surface and mitigate exploration biases.}

\begin{figure}[h!]
    \centering
    \includegraphics[alt={Figure showing availability bias}, width=\linewidth]{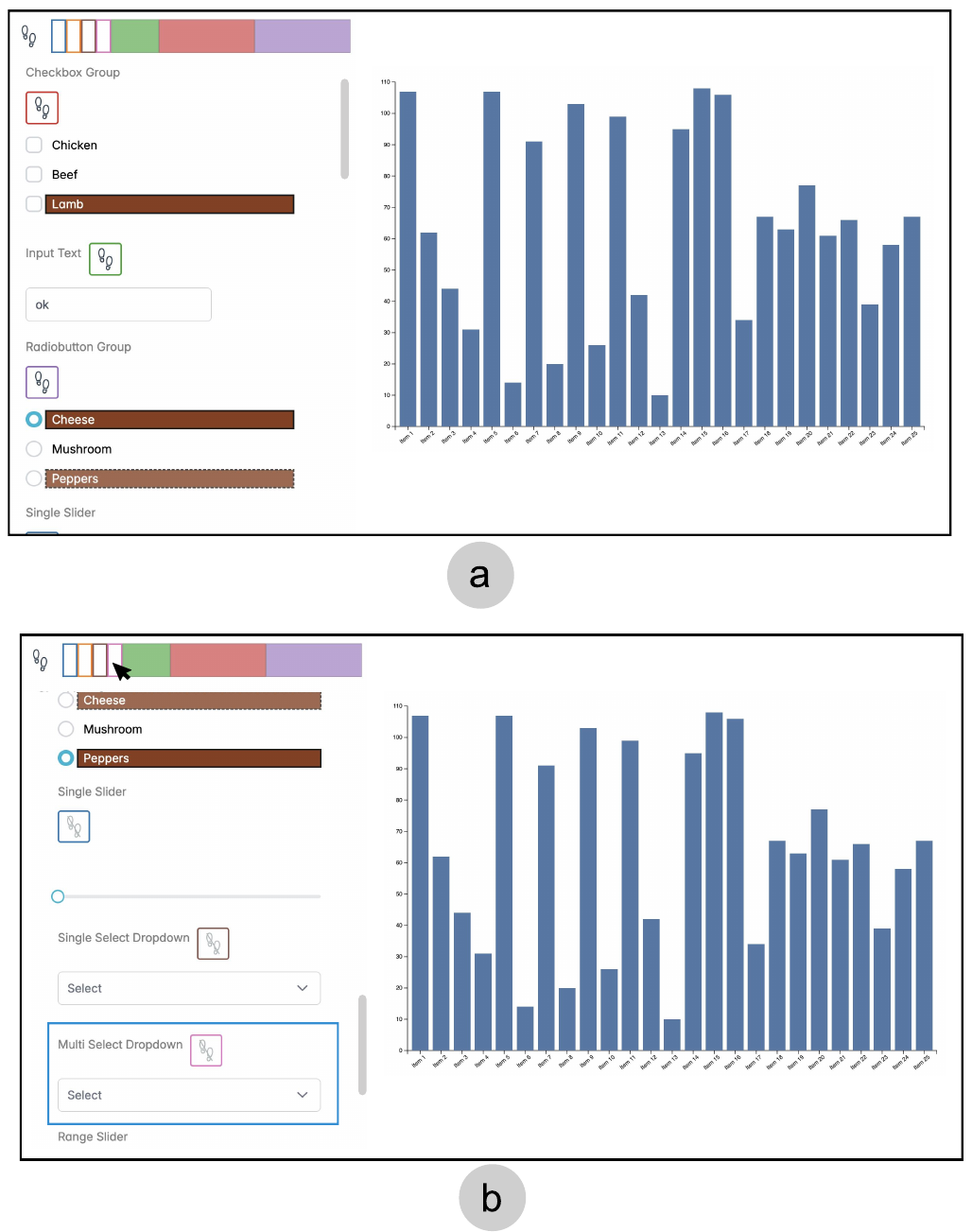}
    \caption{Scenario: (a) Widgets are out of viewport during a visual analysis session; (b) user navigates to specific widgets via the SuperWidget.}
    \label{fig:scenario1}
\end{figure}

Consider a visual analyst exploring a large dataset, such as sales records filtered by demographics and regions, but unintentionally fixating on familiar subsets due to viewport constraints or habitual control use. In traditional systems lacking cross-control summaries, this availability bias goes unnoticed, leading to skewed insights—like overlooking underrepresented regions or demographics—and potentially propagating errors into reports or models.

SuperProvenanceWidgets can empower users to detect such biases at a glance via the SuperWidget. Empty-filled boxes in the Aggregate View (Figure~\ref{fig:scenario1}a) highlight untouched controls, such as out-of-viewport filters for gender or region; clicking them navigates directly to those widgets (Figure~\ref{fig:scenario1}b), prompting balanced exploration. Developers can further leverage the API to reorder widgets by usage frequency, surfacing underused ones prominently; for instance:

\begin{lstlisting}[style=TS, aboveskip=2pt, belowskip=2pt]
const [registeredComponents, setRegisteredComponents] = useProvenance()
for (const [widgetId, widgetData] of registeredComponents) {
  // access widget-level data
}
\end{lstlisting}

\vspace{0.1cm}

This is particularly vital for real-world issues like gender bias, where users might filter to a single value without exploring alternatives, as the Aggregate View provides instant awareness of uneven interaction patterns across the full set of controls.

\subsection{Facilitate user interface design and personalization.}

\begin{figure}[h!]
        \centering
        \includegraphics[alt={High-level process diagram for offline UI layout customization}, width=1\linewidth]{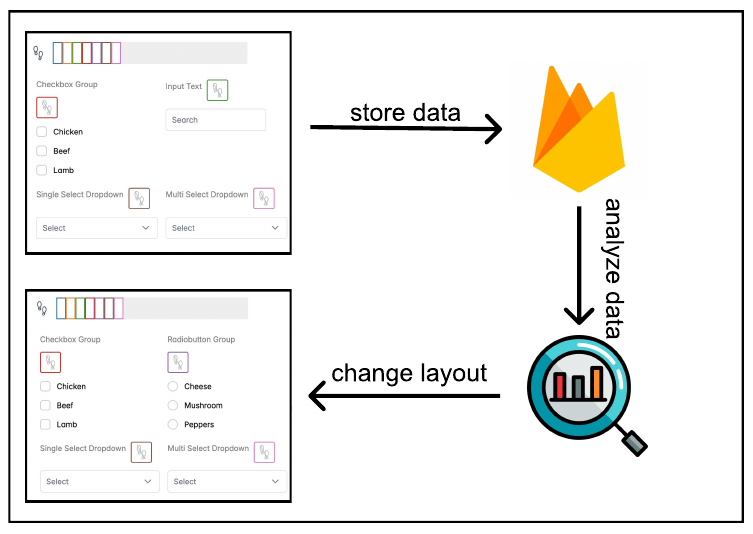}
        \caption{High-level process diagram for offline UI layout customization.}
        \label{fig:architecture}
\end{figure}

Picture a dashboard developer monitoring user sessions on an interactive analytics tool, noticing frustrating patterns like repeated scrolling or tab-switching between interdependent controls--such as color sliders paired with category checkboxes--for trend comparisons. Absent cross-control provenance, these inefficiencies remain hidden, resulting in suboptimal layouts that hinder productivity and user satisfaction.

The SuperWidget uncovers these patterns: frequent co-interactions appear as clustered color pairings in the Aggregate View and tight Gantt bar groups in the Temporal View. Developers can then implement dynamic personalization, auto-grouping high-co-use controls or bubbling low-usage ones to the top—using data streamed to Firebase for team analysis or processed in-memory for real-time tweaks (Figure~\ref{fig:architecture}). Options include manual weekly reviews or automatic in-session reordering, streamlining access and enhancing workflows for all users.
\section{Evaluation: Cognitive Dimensions of Notation}

We present a technical self-assessment of \app using the Cognitive Dimensions of Notations~\cite{blackwell2001cognitive}.

\paragraphHeadingSpace\bpstart{Consistency:} \app exposes a common set of UI components, which behave consistently (described in Section~\ref{ext-design}). The notation is also consistent with the framework it has been implemented in (i.e. React.js), as well as being independent of base libraries (via Web Components).

\paragraphHeadingSpace\bpstart{Viscosity:} The system’s viscosity increases with lower abstraction. The \textbf{compute} module, being more primitive, requires more effort to work with than the \textbf{UI} module.

\paragraphHeadingSpace\bpstart{Abstraction Gradient:} Following goal \textbf{G5}, the library spans multiple abstraction levels, from low-level \textbf{compute} to high-level \textbf{UI} components, enabling developers to choose the appropriate layer for their needs. The abstraction gradient flows from simple use-cases requiring least effort, or low viscosity, to more complex use-cases that span the full function of the library.
\section{Limitations and Future Work}

SuperProvenanceWidgets represents a significant step forward in web-based analytic provenance by enabling developers to track and visualize interactions across multiple UI controls in real-time. Through its modular architecture and innovative SuperWidget, the library addresses key challenges in analysis transparency, bias mitigation, and interface optimization, as demonstrated in our usage scenarios. However, several limitations present opportunities for future enhancement.

\paragraphHeadingSpace\bpstart{Limitations.} While SuperProvenanceWidgets excels at capturing fine-grained interaction data within web-based UI controls, its current implementation has notable constraints. First, it focuses exclusively on client-side provenance tracking in JavaScript environments; it does not natively support server-side computations or integration with backend analytics pipelines, limiting its utility in full-stack applications. Second, the library assumes relatively low-to-moderate interaction volumes—high-frequency sessions with hundreds of rapid interactions may lead to visual clutter in the Temporal View's Gantt chart, despite filtering options. Third, while framework-agnostic via Web Components, adoption may require additional effort for non-React developers unfamiliar with its TypeScript-based APIs. Finally, provenance data remains in-memory by default, raising concerns for long sessions without explicit export mechanisms to persistent storage.

\paragraphHeadingSpace\bpstart{Future Work.} Building on this foundation, several promising directions emerge. We plan to extend provenance capture to hybrid environments, integrating with server-side frameworks like Node.js or Python backends for comprehensive full-stack tracking. Adaptive visualization techniques—such as automatic clustering, zooming, or AI-driven summarization—could mitigate clutter in dense interaction histories. Real-time collaboration features, inspired by tools like Trrack, would enable shared SuperWidgets for team-based auditing. Additionally, leveraging the tracked provenance for proactive guidance, such as subtle nudges toward underused controls or bias alerts (e.g., extending prior work on analytic behavior awareness~\cite{narechania2021lumos}), could transform passive tracking into active support for unbiased exploration. Finally, empirical user studies will validate the library's impact on real-world tasks, measuring outcomes like analysis reproducibility, bias reduction, and productivity gains.

\section{Conclusion}
SuperProvenanceWidgets is a JavaScript library of UI controls that tracks and visualizes users' analytic provenance within and across UI controls, empowering developers to build more transparent and adaptive guidance-enriched visual analytics tools~\cite{narechania2024dissertation}. It extends its predecessor library, ProvenanceWidgets, and is available as open-source software at \textbf{\href{https://ProvenanceWidgets.github.io}{ProvenanceWidgets.github.io}}.

\bibliographystyle{ACM-Reference-Format}
\bibliography{main}

\end{document}